\documentclass[twocolumn,showpacs,preprintnumbers,amsmath,amssymb,nofootinbib]{revtex4}
\usepackage{graphicx}
\usepackage{dcolumn}
\usepackage{bm}
\usepackage{mathrsfs}
\usepackage[dvips]{color}

\begin{document}


\title{\color[rgb]{0.2,0,0.8}
Machian strings as an alternative to dark matter}

\author{David W. Essex}
 \email{D.W.Essex@damtp.cam.ac.uk}
 \affiliation{Centre for Mathematical Sciences, Wilberforce Road, Cambridge, CB3 0WA, England}
\date{\today}

\begin{abstract}
\color[rgb]{0.2,0,0.8}
Dark matter effects may be attributed to interactions between the Machian strings connecting every pair of elementary particles in the observable Universe. A simple model for the interaction between Machian strings is proposed. In the early Universe, the Machian strings of a density perturbation had a spherically symmetric distribution and the interaction with the Machian strings of a test particle is found to give a multiple of the Newtonian gravitational acceleration. In a strong gravitational field, the interaction between Machian strings tends to a constant limit in order to ensure the absence of dark matter effects in the Solar System. Dark matter effects on a galactic scale may be attributed to a change in the distribution of the Machian strings around a galaxy during the process of galaxy formation. The interaction between the Machian strings of a test mass and the Machian strings of a galaxy is considered in detail and the MOND phenomenology for galaxy rotation curves is obtained.
\end{abstract}
\maketitle
\color[rgb]{0.2,0,0.8}

\section{Introduction}
There are two main arguments for the existence of dark matter. Firstly, dark matter is needed on cosmological scales to ensure that the small fluctuations observed in the cosmic microwave background grow sufficiently rapidly to produce galaxies and galaxy clusters~\cite{longair}. Secondly, dark matter is needed on galactic scales to account for the dynamics of galaxy clusters~\cite{zwicky} and the flat rotation curves of spiral galaxies~\cite{rubin}.

It is well known that an excellent fit to the rotation curves of spiral galaxies may be obtained by modifying Newton's law of gravity~\cite{milgrom} and many attempts to account for dark matter effects have since been made by modifying General Relativity, either by adding scalar and vector fields~\cite{teves,mog} or by introducing a more complicated dependence on the metric~\cite{boehmer,mannheim2}. 
The purpose of the present paper is to show that dark matter effects can also be obtained in a completely different way by modifying the model for an elementary particle. 

The conventional model for an elementary particle, such as a quark or an electron, is a dimensionless point. In the Machian string model described in Section~\ref{msm}, an elementary particle consists of a pointlike centre connected by Machian strings to the centres of all the other elementary particles in the observable Universe. The energy in the  Machian strings connecting two masses gives the Newtonian gravitational acceleration and the interaction between Machian strings gives the additional gravitational acceleration usually attributed to dark matter.

It is shown in Section~\ref{growth} that cosmological dark matter is a natural consequence of the interaction between the Machian strings of an overdense region and the Machian strings of a test mass. The additional acceleration on the test mass is a multiple of the Newtonian gravitational acceleration and directed towards the centre of mass of the Machian strings of the overdense region. 

Galaxy formation is discussed in Section~\ref{form}. Before separation from the Hubble flow, the strings of an overdense region had a spherically symmetric distribution. As the region collapses and starts to rotate to form a disc galaxy, the distribution of the strings becomes concentrated in the plane of the galaxy and aquires a cylindrical symmetry. If the strings are assumed to be concentrated in a cylindrical region with thickness proportional to the radius of the galaxy then the form of the additional acceleration changes from $M/r^2$ to $\sqrt M/r$ and the phenomenology of modified Newtonian dynamics (MOND) may be explained. 

Galaxy clusters are discussed in Section~\ref{cluster}. Most of the baryonic mass in a galaxy cluster is in the form of very hot diffuse gas between the galaxies. During a collision between two galaxy clusters, the galaxy and gas distributions become separated since the galaxies pass through one another whereas the the two gas densities interact. Gravitational lensing studies suggest that the gravitational mass density is usually centred on the galaxies but can also be centred on the gas, which is problematic for both modified gravity and for the conventional dark matter model. There is no such difficulty in the string model because the location the gravitational mass density is determined by the centre of mass of the Machian strings and the Machian strings may be centred either on the galaxies or on the gas.

The magnitude of additional acceleration effects in the Solar System is discussed in Section~\ref{solar} and the additional acceleration in the string model is found to be consistent with the experimental limit. The conclusions are presented in Section~\ref{conclude}. 

\section{The Machian string model}\label{msm}
The basic postulate of the Machian string model is that the total energy in all the Machian strings of a particle of rest mass $m$ is constant and equal to $mc^2$. The entire rest mass of a massive particle is distributed in the Machian strings connecting it to other particles. The accompanying paper on dark energy~\cite{paper3} shows that if the energy in a Machian string of length $r$ joining masses $m$ and $M$ has the form
\begin{eqnarray}\label{estring}
 \frac{GmM}{r}\Big(1+\frac{Hr}{2c}\Big)\,,
\end{eqnarray}
where $H$ is the Hubble parameter, then the expansion history of the Universe in the conventional $\Lambda$CDM model is reproduced almost exactly. The string energy consists of positive Newtonian potential energy $GmM/r$ and an additional energy proportional to $H$. The additional energy, which is responsible for the accelerating expansion of the Universe, is independent of the length of the string. 

The present paper is concerned with the force exerted on the centre of a test mass by the Machian strings connected to it. Since there is no force associated with a constant energy, there is no contribution to the force on the centre from the term in~(\ref{estring}) proportional to $H$. For the remainder of the paper, the energy in the strings connecting two masses $m$ and $M$ will therefore be taken to be $GmM/r$. 

\subsection{Newtonian gravity}\label{apptension}
Since the total energy in all the strings of $m$ is equal to $mc^2$, the energy in the strings of $m$ other than those conencted to $M$ is $mc^2-GmM/r$. Similarly, the energy in the strings of $M$ other than those connected to $m$ is $Mc^2-GmM/r$. The total energy in all the strings connected to the two masses is $(mc^2-GmM/r)+(Mc^2-GmM/r)+GmM/r =mc^2 + Mc^2 - GmM/r$ and the interaction energy is therefore the same as in Newtonian gravity. Although Machian strings have positive Newtonian potential energy, all the strings are in tension.

\subsection{The interaction between Machian strings}
If the Machian strings connecting the centre of a test mass to the centres of all the other particles in the observable Universe have a spherically symmetric distribution, the tension forces exerted by the Machian strings on the centre all cancel out. The only force acting on the test mass is then the Newtonian gravitational force due to nearby masses. If there is an interaction between the strings of the test mass and the strings of a mass $M$, however, the strings around the test mass become distorted. The tension forces acting on the centre no longer cancel out and there is an additional force on the centre. The additional force gives the additional gravitational acceleration usually attributed to dark matter.

The simplest assumption is that the interaction is a function of the dimensionless ratio of the density of Machian strings of $M$ to the density of background strings. 
The paths of the strings connected to a test mass and the corresponding string tensions in the presence of a mass $M$  are calculated as described in Appendix~\ref{appmodel}. 
In contrast to the Newtonian gravitational acceleration, which is directed towards the centre of $M$, the additional gravitational acceleration due to the interaction between Machian strings is directed towards the centre of mass of the Machian strings of $M$. 

The form of the additional acceleration depends on the distribution of strings around $M$. Two different distributions will be considered, namely a spherically symmetric distribution and a cylindrically symmetric distribution. 
 


\subsection{The density of strings around $M$}\label{density}
\subsubsection{Spherical distribution}
If the distribution of strings around $M$ is spherically symmetric then, as shown in Appendix~\ref{appdensity},   the ratio of the density of strings around $M$ to the density of background strings is given by 
\begin{eqnarray}\label{urs}
 u\,=\,\frac{M}{r^2}\Big/\frac{M_U}{R_U^2}\,,
\end{eqnarray}
where $M_U$ and $R_U$ are the mass and radius of the observable Universe, respectively. If $\rho$ denotes the spherical polar coordinate radial distance in units of $R_U\sqrt{M/M_U}$, so that $r= \rho R_U\sqrt{M/M_U}$, then
\begin{eqnarray}\label{u1}
u\,=\,\frac{1}{\rho^2}\,.
\end{eqnarray}

\subsubsection{Cylindrical distribution}
Consider a disc galaxy of mass $M$ and radius $R_g$ and suppose the strings around the galaxy have a cylindrically symmetric distribution with an axis of symmetry along the axis of rotation of the galaxy. The precise form of the string distribution depends on the details of the galaxy formation process but it is reasonable to assume that the string density is concentrated near to the plane of the galaxy and decreases with a length scale proportional to the radius of the galaxy, $R_g$. Suppose, therefore, that the string density is proportional to $(z^2+\lambda^2 R_g^2)^{-1}$, where $z$ is the distance from the plane of the galaxy and $\lambda$ is a constant of order unity. Appendix~\ref{appdensity} shows that the ratio of the density of strings around the galaxy to the density of background strings is then
\begin{eqnarray}\label{u3}
u\,=\,\frac{2M\lambda R_g}{\pi r (z^2+\lambda^2R_g^2)}\Big/\frac{M_U}{R_U^2}\,,
\end{eqnarray}
where $r$ now denotes the radial distance in cylindrical polar coordinates. 


It turns out that the radius $R_g$ of a galaxy of mass $M$ is roughly equal to $R_U\sqrt{M/M_U}$. Studies of galaxy rotation curves have shown that there is a transition from the Newtonian acceleration proportional to $M/r^2$ to an acceleration proportional to $\sqrt M/r$ at an acceleration scale $a_0=1.2\times 
10^{-10}$m/s$^2$~\cite{mcgaugh}. Moreover, the transition from a Newtonian rotation curve to a flat rotation curve occurs at approximately $R_g$, the radius of the visible galaxy~\cite{sanders2}. It follows that $R_g$ is given approximately by the equation $GM/R_g^2=a_0$, i.e.
\begin{eqnarray}\label{rg1}
 R_g\,=\,\sqrt{\frac{GM}{a_0}}\,,
\end{eqnarray}
which may be written in the form
\begin{eqnarray}\label{rg}
 R_g\,=\,R_U\sqrt{\frac{M}{M_U}}\sqrt{\Big(\frac{GM_U}{R_Uc^2}\Big)\Big(\frac{c^2}{a_0R_U}\Big)}\,.
\end{eqnarray}
The accompanying paper on dark energy~\cite{paper3} gives 
\begin{eqnarray}\label{gmru}
\frac{GM_U}{R_Uc^2}\,=\,0.30
\end{eqnarray}
and the acceleration scale associated with the expansion of the Universe is
\begin{eqnarray}\label{c2ra}
\frac{c^2}{R_U}\,=\,1.9\times 10^{-10}\,{\mbox m}/{\mbox s}^2\,=\,1.6\,a_0\,.
\end{eqnarray}
Substituting~(\ref{gmru}) and~(\ref{c2ra}) into equation~(\ref{rg}) then gives
\begin{eqnarray}\label{rg2}
R_g\,=\,0.69 \,R_U\sqrt{\frac{M}{M_U}}\,.
\end{eqnarray}
The density ratio~(\ref{u3}) within a distance $\lambda R_g$ from the plane of the galaxy reduces to
\begin{eqnarray}\label{u4}
u\,=\,\frac{2\widetilde\lambda}{\pi\rho(z^2+\widetilde\lambda^2)}\,,
\end{eqnarray}
 where $\widetilde\lambda=0.69\lambda$, $\rho$ is the cylindrical polar coordinate radial distance in units of $R_U\sqrt{M/M_U}$ and $z$ is also in units of $R_U\sqrt{M/M_U}$.


\section{Machian strings and the growth of structure}\label{growth}
\subsection{String interactions in the early Universe}
The density of strings associated with a density fluctuation in the early Universe was very small compared to the background string density. Before the perturbations became large enough to collapse and form bound structures, the strings associated with a given overdense region had a spherically symmetric distribution and the string density ratio $u$ was therefore given by equation~(\ref{u1}). The density ratio increases with time as the perturbations grow but was still less than about $10^{-3}$ at the time of collapse, as shown in Appendix~\ref{appearly}.

The interaction between Machian strings is given by the function $f(u)$, defined in Appendix~\ref{appmodel} as the fractional increase in mass per unit length in the Machian strings of a test mass due to the presence of a mass $M$. Since $u\ll 1$, any analytic function $f(u)$ may be approximated by a linear function so that
\begin{eqnarray}\label{flin}
f(u)\,\approx\,\gamma u
\end{eqnarray}
for some constant $\gamma$. 


\subsection{The effective dark matter density}
The additional gravitational acceleration corresponding to the density ratio~(\ref{u1}) and the linear interaction function~(\ref{flin}) can be calculated analytically for $\rho\gg 1$, as shown in Appendix~\ref{appcalc}, with the result
\begin{eqnarray}\label{edm}
a\,=\,\frac{3\pi^2\gamma}{4\rho^2}\Big(\frac{GM_U}{R_Uc^2}\Big)\frac{c^2}{R_U}\,.
\end{eqnarray}
The acceleration~(\ref{edm}) is centred on the centre of mass of the strings of the overdensity and is a multiple of the Newtonian gravitational acceleration. Indeed, the Newtonian acceleration may be written in terms of $\rho$ in the form
\begin{eqnarray}\label{an}
a_{\text{\tiny N}}\,=\,\frac{GM}{r^2}\,=\,\frac{1}{\rho^2}\Big(\frac{GM_U}{R_Uc^2}\Big)\,\frac{c^2}{R_U}
\end{eqnarray}
so the additional acceleration~(\ref{edm}) corresponds an effective dark matter density, $\rho_{\text{\tiny DM}}$, that is larger than the baryon density, $\rho_b$, by a factor
\begin{eqnarray}
\frac{\rho_{\text{\tiny DM}}}{\rho_b}\,=\,\frac{3\pi^2\gamma}{4}\,.
\end{eqnarray}
Analysis of the CMB and matter power spectra in the conventional $\Lambda$CDM model show that the ratio of dark matter to baryonic matter is the early Universe is $\rho_{\text{\tiny DM}}/\rho_b=\Omega_c/\Omega_b=5.0$~\cite{bennett}. The corresponding value of $\gamma$ in the string model is
\begin{eqnarray}\label{alpha2}
\gamma\,=\,\frac{20}{3\pi^2}\,=\,0.68\,.
\end{eqnarray}

The time evolution of perturbations in the string model is identical to the time evolution in $\Lambda$CDM cosmology until either the values of $u$ are such that the linear approximation~(\ref{flin}) no longer applies or the distributions of strings are no longer spherically symmetric. The spherical symmetry of the strings is expected to break down during the process of galaxy formation, as discussed in Section~\ref{form}.


\section{Galaxy formation}\label{form}
%

\subsection{Gravitational collapse}\label{gc}
As in conventional cosmology, particles are gravitationally attracted towards an overdense region and the matter density of the overdense region eventually becomes much larger than the background matter density. The region then stops expanding with the Hubble flow and undergoes gravitational collapse to form stars and galaxies~\cite{longair2}. It is shown in Appendix~\ref{appearly} that an overdense region separating from the Hubble flow at $z\sim 3$ decreases in size by a factor $\sim 35$ during the galaxy formation process.

The large decrease in radius during the collapse phase leads to a large increase in the rotational velocity of the infalling matter due to the conservation of angular momentum~\cite{fall}. In the string model, the rotation causes the strings connected to the infalling matter to acquire the same cylindrical symmetry as the resulting disc galaxy.

\subsection{Transition to the MOND regime}\label{trans}
It is well known that an excellent fit to the observed galaxy rotation curves is obtained if the Newton gravitational acceleration $a_N=GM/r^2$ is replaced by the MOND~\cite{milgrom} acceleration $a_M$ defined by
\begin{eqnarray}\label{am}
a_M\,=\,
\left\{
	\begin{array}{ll}
		\hspace*{0.5cm}a_N & \hspace*{0.4cm}\mbox{if}\hspace*{0.4cm} a_N \ge a_0\\
		\sqrt{a_Na_0} & \hspace*{0.4cm}\mbox{if}\hspace*{0.4cm} a_N < a_0\,.
	\end{array}
\right.
\end{eqnarray}
With the galaxy radius $R_g$ defined by equation~(\ref{rg1}), the transition from the Newtonian acceleration to the acceleration $\sqrt{a_Na_0}$ occurs at exactly $r=R_g$. The rotation curve $v(r)=\sqrt{ra(r)}$ is proportional to $1/\sqrt r$ in the Newtonian regime, $r<R_g$, and has a constant value $(GMa_0)^{1/4}=\sqrt{a_0R_g}$ in the MOND regime $r>R_g$. 

\begin{figure}[h]
\hspace*{-3cm}
\includegraphics[height=6cm,width=5cm]{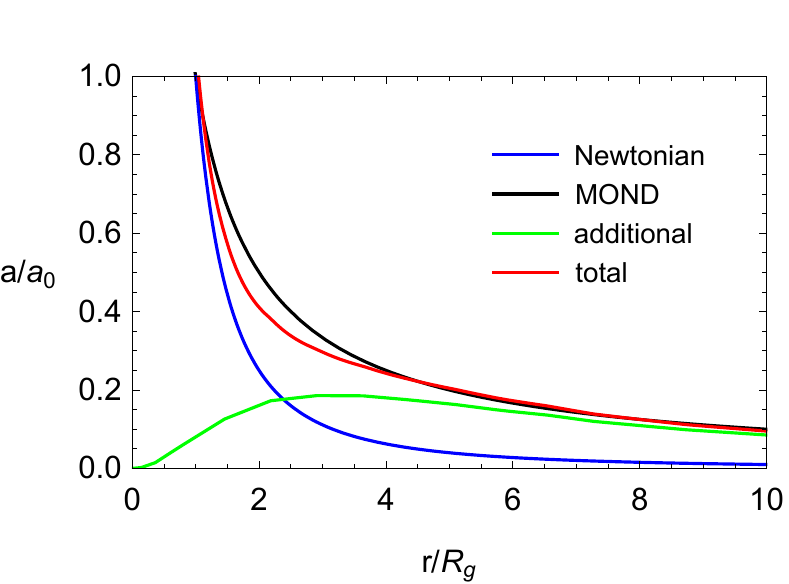}
\caption{\label{ctotal} The acceleration on a test mass in the plane of a disc galaxy of radius $R_g$ when the strings of the galaxy have the cylindrical distribution given by equation~(\ref{u4}) with $\lambda=0.5$. The interaction between the Machian strings of the test mass and the Machian strings of the galaxy is defined by the function~(\ref{intf2}) of Appendix~\ref{appmodel} with $\alpha=0.68$ and $\beta=1$ and the additional acceleration (green) is calculated using equation~(\ref{af}) of Appendix~\ref{appcalc}. The additional acceleration is added to the Newtonian acceleration (blue) given by~(\ref{an}) to calculate the total acceleration (red). The MOND acceleration given by~(\ref{am}) is also shown.}
\end{figure}

\begin{figure}[h]
\hspace*{-3.5cm}
\includegraphics[height=6cm,width=5cm]{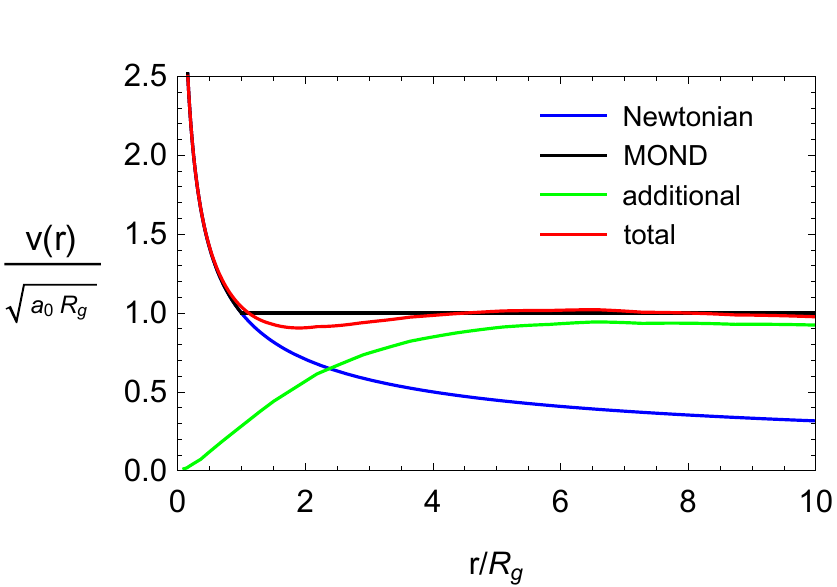}
\caption{\label{rotation} The rotation curves corresponding to the accelerations in Figure~\ref{ctotal}. The rotation curve for the total acceleration in the string model is very close to the MOND rotation curve.}
\end{figure}

The additional acceleration in the string model was calculated numerically\footnote{The Mathematica~\cite{mathematica} code is available upon request.}, as described in Appendix~\ref{appcalc}, for a cylindrical string distribution with the density of strings around the galaxy given by equation~(\ref{u4}). The total gravitational acceleration was then calculated by adding the additional acceleration to the Newtonian acceleration and compared with the MOND acceleration~(\ref{am}).

A close fit to the MOND acceleration was found for $\alpha=0.68$, $\beta=1$ and $\lambda=0.5$, as shown in Figure~\ref{ctotal}. The corresponding rotation curves are shown in Figure~\ref{rotation}. The value of $\beta$ has only a small effect on the rotation curve and the effect of varying the parameters $\alpha$ and $\lambda$ is discussed in Appendix~\ref{appcalc}. 


\section{Galaxy clusters}\label{cluster}
The gravitational mass distribution in an isolated galaxy cluster is not consistent with simple modified gravity theories such as MOND but can be accounted for using the additional acceleration in the string model~\cite{paper2n}. 
The case of colliding galaxy clusters, in which the galaxy and gas distributions become separated, is even more challenging for modified gravity theories. In a modified gravity theory the gravitational mass density should be centred on the most massive component, namely the gas, since a modification of the law of gravity increases the strength of the gravitational field but does not change the centre of gravitational attraction. In the Bullet Cluster~\cite{clowe}, however, the gravitational mass density is centred on the galaxies. The result is consistent with the conventional model of collisionless dark matter because the dark matter and the galaxies in one cluster simply pass through the dark matter and the galaxies in the other cluster and therefore remain together. The result is also consistent with the string model since the location of the gravitational mass density is determined by the centre of mass of the strings and the strings also pass through each other since they are not charged. Almost the entire length of the strings are unaffected by the collision, apart from small sections at the ends of the strings connected to the gas particle centres, so the centre of mass of the strings remains centred on the galaxies.

In some clusters, such as Abell\,520~\cite{jee}, there is evidence that the gravitational mass is centred on the gas rather than the galaxies, contrary to the prediction of the conventional dark matter model. In the string model, the strings of the gas particles are distorted during a collision because the ends are connected to the gas particle centres. The strings will eventually straighten out, however, since the strings are in tension. Moreover, the positions of the gas particle centres are unaffected by the straightening of the strings since the force exerted by the strings on the centres are negligible compared to the electromagnetic forces. Indeed, the typical magnetic field in a galaxy cluster is of order $10^{-10}$\,T~\cite{govoni} and the magnetic force on an electron moving at speed $7\times 10^7$\,m/s, corresponding to a temperature of 10$^8$\,K, gives an acceleration of order 10$^9$\,m/s$^2$ which is very much larger than the maximum acceleration of order $c^2/R_U$ exerted by the strings. When the strings straighten out, the centre of mass of the strings returns to the centre of mass of the gas. It is therefore possible for the gravitational mass to be centred on either the galaxies or the gas in the string model.

\section{The Solar System}\label{solar}
The absence of dark matter effects in the Solar System implies that any additional acceleration is less than $7.5\times 10^{-5}\,a_0$ at the radius of Saturn~\cite{pitjev}. Since the density of Machian strings of the Sun completely dominates the density of Machian strings of the galaxy and the density of background Machian strings, the strings around the Sun have a spherically symmetric distribution. Taking the radius of Saturn's orbit to be $1.5\times 10^{12}\,$m gives $M/r^2\approx 8.9\times 10^5$\,kg/m$^2$ and the values $M_U=9.3\times 10^{22}$ Solar masses and $R_U=4.6\times 10^{26}\,$m from~\cite{paper3} give $M_U/R_U^2=0.88$, so the density ratio~(\ref{urs}) is $u\approx 1.0\times 10^6$. The corresponding value of $\rho$ at the radius of Saturn is $\rho\approx 1.0\times 10^{-3}$.

\begin{figure}[h]
\hspace*{-3.5cm}
\includegraphics[height=6cm,width=5cm]{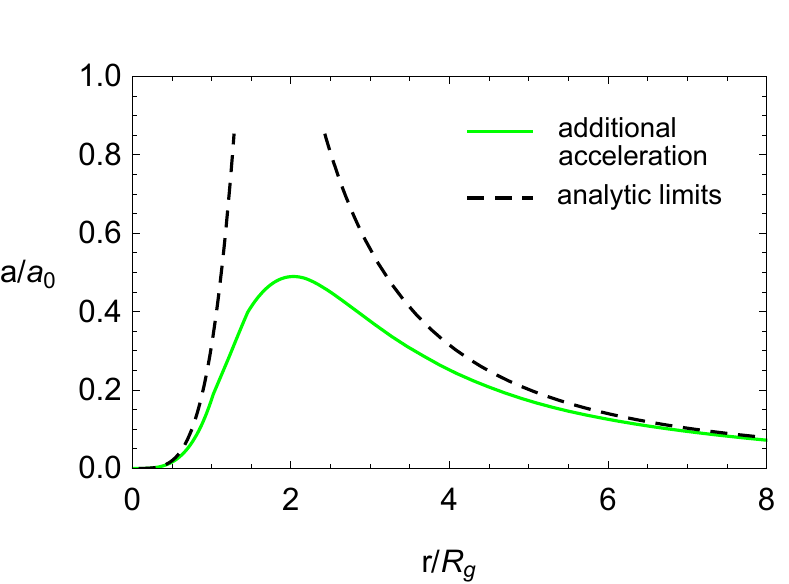}
\caption{\label{spherical} The additional acceleration for a spherically symmetric string distribution when $\alpha=0.68$ and $\beta=1$. The dashed lines show the analytic limits for $\rho\gg 1$ and $\rho\ll 1$ given by equations~(\ref{edm}) and~(\ref{limit}), respectively. Note that, from equation~(\ref{rg2}), $\rho=0.69 \,r/R_g$.}
\end{figure}

The additional acceleration for a spherically symmetric string distribution was calculated numerically as described in Appendix~\ref{appcalc}. The result, for the same parameters as in Section~\ref{form}, is shown in Figure~\ref{spherical}. 
In the limit $\rho\rightarrow \infty$, the numerical result agrees with the analytic result for a spherical string distribution given previously in equation~(\ref{edm}). The additional acceleration in the limit $\rho\rightarrow 0$ 
is given by
\begin{eqnarray}\label{limit}
a\,\approx\, \frac{3\rho^4}{\beta}\Big(\frac{GM_U}{R_Uc^2}\Big)\frac{c^2}{R_U}\,=\,\frac{0.9\rho^4}{\beta}\frac{c^2}{R_U}\,,
\end{eqnarray}
as explained in Appendix~\ref{appcalc}. For $\rho\approx 1.0\times 10^{-3}$, the additional acceleration with $\beta=1$ is about $1.4\times 10^{-12}a_0$, which is well below the experimental limit.


\section{Conclusion}\label{conclude}
The solution to the dark matter problem may require the conventional model of an elementary particle as a dimensionless point to be revised. In the Machian string model, the entire mass energy of an elementary particle is in the Machian strings connecting the centre of the particle to the centres of all the other elementary particles in the observable Universe. A cosmological dark matter density arises naturally from the interaction between Machian strings and the MOND phenomenology required to account for the observed galaxy rotation curves may be attributed to a change in the distribution of Machian strings during the process of galaxy formation.
 
\section{Acknowledgements}
The hospitality of Mr Robert Buis and Mrs Joy Buis in Wartburg, South Africa, is gratefully acknowledged.

\begin{appendix}
 \section{The Machian strings of a test mass}\label{appmodel}
\subsection{The interaction between Machian strings}\label{appintn}
Consider one of the Machian strings of a test mass $m$ connected to a distant mass $\overline m$. In the absence of any interactions between the Machian strings, the string is straight and has total energy $Gm\overline m/R$, where $R$ is the length of the string. The energy per unit length is $T(R)= Gm\overline m/R^2$, assuming the energy to be distributed uniformly along the string. The interaction between Machian strings is specified by an interaction function $f$ that gives the fractional increase in mass per unit length in the Machian strings of a test mass due to the presence of a mass $M$.
 The function $f$ is assumed to be a function of $u$, the ratio of the density of strings of $M$ to the background density of strings, so the energy per unit length of the Machian strings of the test mass $m$ at the point ${\bf x}$ may be written in the form
\begin{eqnarray}\label{dens}
 {\mathcal E}(R,{\bf x})\,=\,T(R)\Big\{1 + f[u({\bf x})]\Big\}\,.
\end{eqnarray}

\subsection{The interaction function}\label{appf}
It follows from the definition~(\ref{dens}) that the interaction function $f(u)$ tends to zero as $u$ tends to zero because the interaction is due to the strings and the density of strings tends to zero as $u$ tends to zero. The requirement that there is no detectable dark matter in the Solar System can be satisfied if the function $f(u)$ saturates sufficiently rapidly at some constant value in the limit $u\rightarrow \infty$. For large values of $u$, the interaction function is then approximately uniform and a uniform change in the energy of the strings of a test mass produces no additional acceleration, by symmetry. The simplest such function has the form
\begin{eqnarray}\label{intf2}
f(u)\,=\,\frac{A(\alpha u+\beta u^2)}{1 + \alpha u + \beta u^2}\,,
\end{eqnarray}
where $A$, $\alpha$ and $\beta$ are free parameters. The function~(\ref{intf2}) vanishes at $u=0$ and tends rapidly to the constant $A$ as $u\rightarrow\infty$. 

In the limit $u\ll 1$, $f(u)\approx \gamma u$ where
\begin{eqnarray}\label{alpha}
\gamma\,=\,A\alpha\,. 
\end{eqnarray}
The value of $\gamma$ needed to give the required cosmological dark matter density is given in equation~(\ref{alpha2}). For a given value of $\alpha$, the corresponding value of $A$ is 
\begin{eqnarray}\label{aa}
A\,=\,\frac{0.68}{\alpha}\,.
\end{eqnarray}


\subsection{String paths and string tension}\label{apppos}
\subsubsection{Variation of total string energy}
 Let $s$ be the path length along one of the Machian strings of a test mass $m$ connected to a distant mass $\overline m$, with $s= s_i$ at $m$ and $s= s_f$ at $\overline m$. If ${\bf x}(s)$ denotes the position of the point along the string at path length $s$ then ${\bf x}(s_i)={\bf X}_i$ and ${\bf x}(s_f)={\bf X}_f$, where ${\bf X}_i$ and ${\bf X}_f$ are the positions of the centres of the masses $m$ and $\overline m$, respectively. The total energy in the string is
\begin{eqnarray}\label{e2}
 E=\int_{s_i}^{s_f}\! {\mathcal E}(R,{\bf x})~ds\,,
\end{eqnarray}
 where ${\mathcal E}(R,{\bf x})$ is the energy per unit length defined in~(\ref{dens}) and $R=s_f-s_i$ is the length of the string. It follows from the basic postulate of the string model stated in Section~\ref{msm} that the total energy of all the strings connected to the masses $m$ and $\overline m$ is $E_S\,=\,mc^2 + \overline mc^2- E$. To minimise the total energy of the system it is therefore necessary to find a string path for which the energy~(\ref{e2}) is a maximum. Since the energy~(\ref{e2}) tends to zero in the limit that the string becomes infinitely long, string paths that maximise~(\ref{e2}) do exist.

 Consider a variation $\delta {\bf x}$ of the string path with the string held fixed at the distant mass
 $\overline m$, so that $\delta{\bf x}= \delta {\bf X}_i$, say, at $s=s_i$ and $\delta{\bf x}= {\bf 0}$ at
 $s=s_f$. Since the total length of the string changes it is convenient to introduce the parameter $\sigma$ along
 the string path so that $\sigma$ is fixed at $m$ and $\overline m$, with $\sigma=0$ at $m$ and $\sigma=1$ at $\overline m$. The energy $E$ is then given by
\begin{eqnarray}\label{t1}
 E=\int_0^1\! {\mathcal E}(R,{\bf x})\,|{\bf x}^\prime|~d\sigma\,,
\end{eqnarray}
 where the prime denotes differentiation with respect to $\sigma$, and the string length $R$ is given by
\begin{eqnarray}\label{r1}
 R=\int_0^1\! |{\bf x}^\prime|~d\sigma\,.
\end{eqnarray}
 The variation of~(\ref{t1}) is
\begin{eqnarray}\label{t2}
 &&  \hspace*{-0.8cm}\delta E= \!\int_0^1\! \Big\{\Big(\frac{\partial{\mathcal E}}{\partial R}\delta R +
 {\pmb\nabla}{\mathcal E}.\delta{\bf x}\Big)\,|{\bf x}^\prime| + {\mathcal E}\frac{{\bf x}^\prime.\delta{\bf
 x}^\prime}{|{\bf x}^\prime|}\Big\}~d\sigma\,,~~ \\ &&  \hspace*{-0.3cm}\mbox{where} \hspace*{1.3cm} \delta R\,=
 \int_0^1\!\frac{{\bf x}^\prime.\delta{\bf x}^\prime}{|{\bf x}^\prime|}~d\sigma\,.\label{r2}
\end{eqnarray}
 Integrating~(\ref{r2}) by parts gives
\begin{eqnarray}\label{r3}
 \delta R= \Big(\frac{{\bf x}^\prime.\delta{\bf x}}{|{\bf x}^\prime|}\Big)_0^1\,-
  \int_0^1\! \Big\{\frac{{\bf x}^{\prime\prime}}{|{\bf x}^\prime|}- \frac{{\bf x}^\prime({\bf x}^\prime.
  {\bf x}^{\prime\prime})}{|{\bf x}^\prime|^3}\Big\}.\delta{\bf x}~d\sigma.~~~~
\end{eqnarray}
 After changing back to the path length parameterisation, for which $|{\bf x}^\prime|=1$ and ${\bf x}^\prime.
 {\bf x}^{\prime\prime}=0$,~(\ref{r3}) becomes
\begin{eqnarray}\label{r4}
 \delta R\,=\, -{\bf x}^\prime(s_i).\delta {\bf X}_i\, -\int_{s_i}^{s_f}\! {\bf x}^{\prime\prime}.\delta{\bf
 x}\,\,ds\,.
\end{eqnarray}
 Similarly,~(\ref{t2}) becomes
\begin{eqnarray}\label{t3}
 && \hspace*{-0.8cm}\delta E\,=\, -\,{\mathcal E}(R,{\bf X}_i)\,{\bf x}^\prime(s_i).\delta {\bf X}_i\,
 \nonumber\\ && \hspace*{-0.8cm}+\int_{s_i}^{s_f}\!\!  \Big\{\frac{\partial{\mathcal E}}{\partial R}\delta R
 + \!\Big[{\pmb\nabla}{\mathcal E} - {\bf x}^\prime ({\bf x}^\prime.{\pmb\nabla}{\mathcal E}) -
 {\mathcal E}{\bf x}^{\prime\prime}\Big].\delta{\bf x}\Big\}\,ds.~\hspace*{-0.5cm}\nonumber\\
\end{eqnarray}
 The variation $\delta R$ is independent of $s$ and may be taken outside the integral. Substituting for
 $\delta R$ from~(\ref{r4}) then gives
\begin{eqnarray}\label{t4}
 && \hspace*{-1cm}\delta E\,=\, -\,\Big[{\mathcal E}(R,{\bf X}_i) + I(R)\Big]{\bf x}^\prime(s_i).\delta {\bf
 X}_i\, \nonumber\\ && \hspace*{-0.8cm}+\int_{s_i}^{s_f} \!\Big\{{\pmb\nabla}{\mathcal E} - {\bf x}^\prime
 ({\bf x}^\prime.{\pmb\nabla}{\mathcal E}) - [{\mathcal E}+I(R)]{\bf x}^{\prime\prime}\Big\}.\delta{\bf x}\,ds,~~~
 \\ && \hspace*{-0.3cm}\mbox{where} \hspace*{1cm} I(R)\,=\, \int_{s_i}^{s_f}\!\frac{\partial{\mathcal E}}
 {\partial R}\,ds\,.\label{ir}
\end{eqnarray}

\subsubsection{The path equation}
 The requirement that $E$ is stationary for all variations of the string path for which the string is fixed at
 both ends, i.e. for which $\delta{\bf X}_i=0$, gives the path equation
\begin{eqnarray}\label{pe}
 {\pmb\nabla}{\mathcal E} - {\bf x}^\prime
 ({\bf x}^\prime.{\pmb\nabla}{\mathcal E}) - [{\mathcal E}+I(R)]{\bf x}^{\prime\prime}\,=\,0\,.
\end{eqnarray}
The component of equation~(\ref{pe}) along the direction of the string, parallel to ${\bf x}^\prime$, is identically zero and the component along a direction ${\bf m}$ perpendicular to the string is
\begin{eqnarray}\label{curv}
 {\bf m}.{\bf x}^{\prime\prime}\,=\,\frac{{\bf m}.{\pmb\nabla}{\mathcal E}}{{\mathcal E}(R,{\bf x})+I(R)}\,.
\end{eqnarray}
Equation~(\ref{curv}) may be integrated numerically to calculate the string paths as described in Appendix~\ref{appcalc}.

\subsubsection{The tension in the strings}
 For paths satisfying the path equation~(\ref{pe}), it follows from~(\ref{t4}) that the change in the total energy of the string when mass $m$ is displaced by $\delta{\bf X}_i$ is
\begin{eqnarray}\label{c1}
 \delta E\,=\, -\,\Big[{\mathcal E}(R,{\bf X}_i) + I(R)\Big]{\bf x}^\prime(s_i).\delta {\bf X}_i\,.
\end{eqnarray}
 The total energy of the system is $E_S\,=\,mc^2 + \overline mc^2- E$, so
\begin{eqnarray}\label{c2}
 \delta E_S\,=\, \Big[{\mathcal E}(R,{\bf X}_i) + I(R)\Big]{\bf x}^\prime(s_i).\delta {\bf X}_i\,.
\end{eqnarray}
 If ${\bf F}$ denotes the force exerted by the string on the mass $m$ then the work done by the system is
 ${\bf F}.\delta{\bf X}_i$, so $\delta E_S\,=\,-{\bf F}.\delta{\bf X}_i$. The force exerted on the mass $m$
 is therefore
\begin{eqnarray}\label{force}
 {\bf F}\,=\, -\Big[{\mathcal E}(R,{\bf X}_i) + I(R)\Big]{\bf x}^\prime(s_i)\,,
\end{eqnarray}
 where ${\bf x}^\prime$ is the unit vector from $m$ to $\overline m$. The string tension at a general point ${\bf x}$ along the string is given by
\begin{eqnarray}\label{tens}
 T(R,{\bf x})\,=\,-[{\mathcal E}(R,{\bf x})+I(R)]\,.
\end{eqnarray}
 Note that, in the absence of any interactions between the strings, ${\mathcal E}= T(R)\,=\,Gm\overline m/R^2$
 and~(\ref{force}) reduces to the Newtonian gravitational force
\begin{eqnarray}
 {\bf F}\,=\, \frac{Gm\overline m}{R^2}\,{\bf x}^\prime\,.
\end{eqnarray}

 Consider the energy per unit length ${\mathcal E}(R,{\bf x})$ defined by equation~(\ref{dens}). After substituting~(\ref{dens}) into~(\ref{ir}), noting that $T(R)$ is proportional to $1/R^2$, the string
 tension~(\ref{tens}) becomes
\begin{eqnarray}\label{ei}
 T(R,{\bf x})\,=\,T(R)\,F(R,{\bf x})\,,
\end{eqnarray}
 say, where
\begin{eqnarray}\label{gu}
 F(R,{\bf x})\,=\,1 - f[u({\bf x})] + \frac{2}{R}\int_{s_i}^{s_f}\!\!f[u({\bf x}(s))]\,ds.~~
\end{eqnarray}
 To ensure that the curvature~(\ref{curv}) remains finite, the string tension must be positive everywhere along
 the string so the function $F(R,[\bf x])$ must be positive. For a string of length $R_U$, the change of variables
 $s=\sigma R_U\sqrt{M/M_U}$ gives
\begin{eqnarray}\label{gu2}
 F(R,{\bf x})&=&1 - f[u({\bf x})] + 2\sqrt{\frac{M}{M_U}}\int_0^{\sqrt{\frac{M_U}{\!M}}}\!\!f(u)
 \,d\sigma\,,\hspace*{-0.5cm}\nonumber\\
&\approx &  1- f[u({\bf x})]\,,
\end{eqnarray}
 where $u=(\rho^2+\sigma^2-2\rho\sigma\cos\theta)^{-1}$, $\rho$ is defined by $r=\rho R_U\sqrt{M/M_U}$ and $\theta$ is the angle between ${\bf x}$ and ${\bf x}_M$. 
It follows that $F(R,{\bf x})<1$, so the interaction with $M$ reduces the tension in the Machian strings of $m$. The condition needed to ensure that the string tension remains positive is 
\begin{eqnarray}\label{tpc}
f(u)\,<\,1\,.
\end{eqnarray}
After substituting~(\ref{gu2}) into~(\ref{ei}), the tension in the strings of $m$ takes the form
\begin{eqnarray}\label{tm}
 T(R,{\bf x})\,=\,T(R)\{1- f[u({\bf x})]\}
\end{eqnarray}
and the tension in the strings at the centre of $m$ is therefore
\begin{eqnarray}\label{tm0}
 T(R)[1- f(u_0)]\,,
\end{eqnarray}
where $u_0$ is the value of $u$ at the centre of $m$.

 \section{The density of strings around the mass M}\label{appdensity}
\subsection{Spherically symmetric distribution}\label{appsph}
Consider first the simple case where all the Machian strings have length $R_U$ and let $\mu$ denote the mass per unit length in the Machian strings of length $R_U$ joining two particles of unit mass. The Machian strings connected to a mass $M$ then contain a mass $\mu r MM_U$ within a distance $r$ of $M$, where $M_U$ is the mass of the observable Universe. If the strings of $M$ have a spherically symmetric distribution, the corresponding string density $\rho_s$ is given by $4\pi \rho_s r^2dr = \mu dr M M_U$ from which it follows that 
\begin{eqnarray}\label{rds}
 \rho_s\,=\, \frac{\mu MM_U}{4\pi r^2}\,.
\end{eqnarray}
The mass density in the background Machian strings, $\rho_0$, is given by $\rho_0=(\mu M_U^2R_U/2)/(4\pi R_U^3/3)$, so $\rho_s/\rho_0 \sim (M/M_U)(R_U^2/r^2)$. When the different lengths of Machian strings are taken into account, the densities $\rho_s$ and $\rho_0$ change by constant factors of order unity. The density ratio $u$, namely the ratio of the density of strings around $M$ to the density of background strings, is therefore equal to $\rho_s/\rho_0$ times a factor of order unity. The constant factor may be absorbed into a redefinition of the parameters $\alpha$ and $\beta$ in the interaction function~(\ref{intf2}) and the density of strings for a spherically symmetric distribution may therefore be defined as
\begin{eqnarray}\label{us}
 u\,=\,\frac{M}{r^2}\Big/\frac{M_U}{R_U^2}\,,
\end{eqnarray}
where $r$ is the distance from the centre of the mass $M$.

\subsection{Cylindrically symmetric distribution}\label{appcyl}
Suppose the mass $\mu dr M M_U$ enclosed in a spherical region of thickness $dr$ becomes redistributed during the process of galaxy formation within a cylindrical region of thickness $dr$ and circumference $2\pi r$, where $r$ now denotes the radial distance in cylindrical polar coordinates. If the new density of strings in the  cylindrical distribution is $\rho_c$ and the mass within the strings is conserved then
\begin{eqnarray}\label{rdc2}
 2\pi rdr\!\int_{-\infty}^\infty\rho_c\,dz\,=\,\mu dr M M_U\,,
\end{eqnarray}
where the $z$ axis is normal to the plane of the galaxy. Suppose, for definiteness, that the density of strings has a $z$-dependence of the form $(z^2+\lambda^2R_g^2)^{-1}$, so that the density is concentrated within a distance of order $\lambda R_g$ from the plane of the galaxy. Equation~(\ref{rdc2}) then gives
\begin{eqnarray}\label{rdc}
 \rho_c\,=\, \frac{\mu MM_U\lambda R_g}{2\pi^2 r(z^2+\lambda^2R_g^2)}\,,
\end{eqnarray}
which is larger than the string density~(\ref{rds}) by a factor 
\begin{eqnarray}\label{rfac}
 \frac{\rho_c}{\rho_s}\,=\,\frac{2r\lambda R_g}{\pi(z^2+\lambda^2R_g^2)}\,.
\end{eqnarray}
The corresponding density ratio $u$ is larger than the density ratio~(\ref{us}) by the same factor, giving
\begin{eqnarray}\label{uc}
u\,=\,\frac{2M\lambda R_g}{\pi r (z^2+\lambda^2R_g^2)}\Big/\frac{M_U}{R_U^2}\,.
\end{eqnarray}

 \section{The density of strings in the early universe}\label{appearly}
Consider an overdense region in the early universe of radius $r(t)$ containing an excess mass $M(t)$. The density of strings of the mass $M(t)$ compared to the background density of strings is given by the ratio $u(t)= (M(t)/r(t)^2)/(M_U(t)/R_U^{\,2}(t))$, where $M_U(t)$ and $R_U(t)$ are the mass and radius of the observable universe at time $t$. In terms of the matter overdensity $\delta(t)\sim (M(t)/r(t)^3)/(M_U(t)/R_U^{\,3}(t))$ the density of strings is given by $u(t)\sim \delta(t)r(t)/R_U(t)$. The radius $R_U(t)$ is proportional to $t$ and $r(t)=Ra(t)$, where $R$ is the comoving radius of the overdensity and $a(t)$ is the scale factor. In the linear regime where $\delta\ll 1$, $R$ is constant since the overdense region expands with the Hubble flow. In the matter era, $\delta(t)$ is proportional to the scale factor $a(t)$ and $a(t)\propto t^{2/3}$. Thus, up to the time at which the perturbations become nonlinear, $u(t)$ increases with time proportional to $a(t)^2/t\propto t^{1/3}$. At the time of nonlinearity, when $\delta\sim 1$, $u(t)\sim r(t)/R_U(t)$. Since $r(t)\ll R_U(t)$ for galaxies and galaxy clusters it is clear that $u\ll 1$ for all physically relevant scales in the linear regime. 

\begin{figure}[h]
\hspace*{-3.4cm}
\includegraphics[height=6cm,width=5cm]{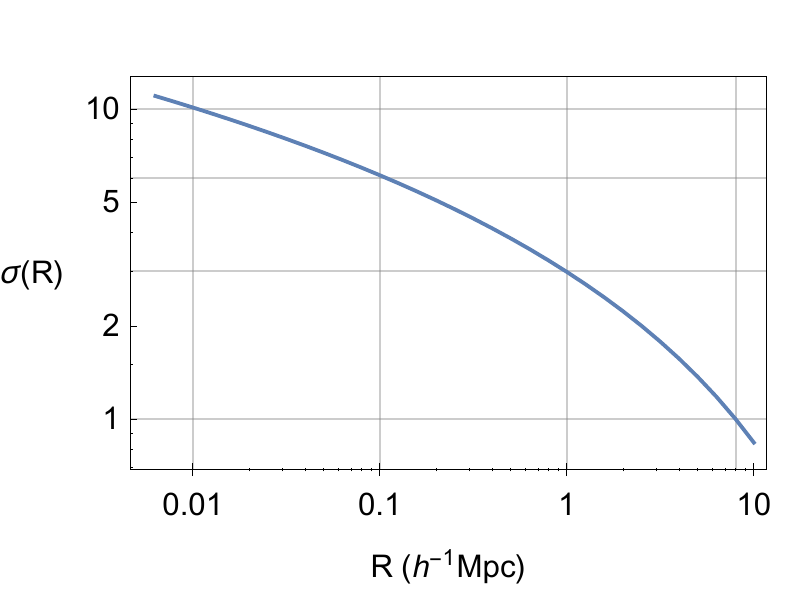}
\caption{\label{sigma} The mass fluctuation in a region of size $R$ at the present time. Regions of size less than $8h^{-1}\,$Mpc have $\sigma > 1$ and have already collapsed to form bound structures.}
\end{figure}

\begin{figure}[h]
\hspace*{-3.4cm}
\includegraphics[height=6cm,width=5cm]{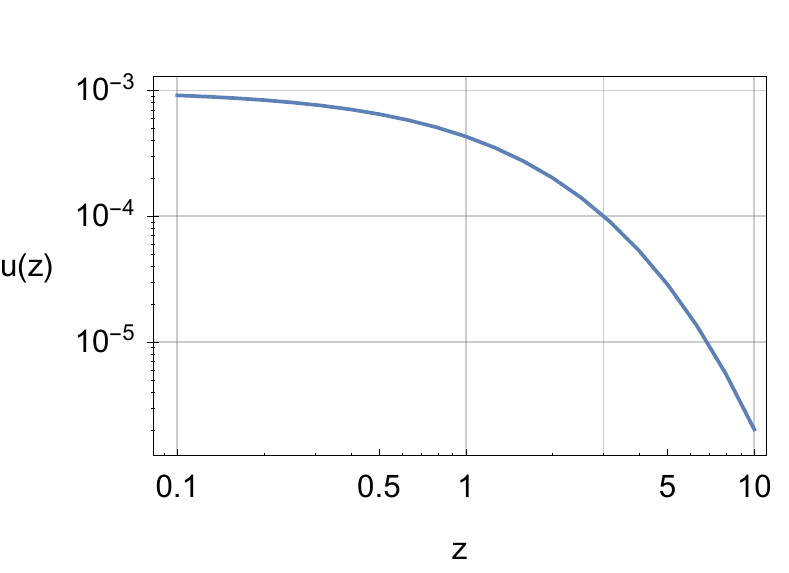}
\caption{\label{uplot} The ratio $u(z)$ of the density of strings to the background density of strings for mass fluctuations becoming nonlinear at redshift $z$.}
\end{figure}

To calculate values of $u$ explicitly it is necessary to find the time at which perturbations on a given comoving scale became nonlinear. If $R(z)$ is the comoving size of an overdense region that becomes nonlinear at redshift $z$ then, at the time of nonlinearity, $r(t)=R(z)/(1+z)$. Since $R_U(t)\propto a(t)^{3/2}$ in the matter era it follows that $R_U(t)=R_U/(1+z)^{3/2}$ so the value of $u$ for an overdense region becoming nonlinear at redshift $z$ was \begin{eqnarray}\label{uz1}
u(z)\,\sim\,\frac{\sqrt{1+z}\,R(z)}{R_U}\,.
\end{eqnarray}
The condition for an overdense region of comoving size $R$ to become nonlinear at redshift $z$ is $\sigma(R,z)\sim 1$, where $\sigma(R,z)$ is the root mean square density fluctuation on a comoving scale $R$ at redshift $z$. Since perturbations grow as $a(t)$ and $a(t)=1/(1+z)$ it follows that $\sigma(R,z)= \sigma(R)/(1+z)$, where $\sigma(R)$ is the root mean square density fluctuation on a comoving scale $R$ at the present time. The function $\sigma(R)$ may be calculated from the observed matter power spectrum~\cite{zentner} and is given approximately by the curve $\sigma(R)= \log_{10}^{\,1.7}(80\,h^{-1}$Mpc$/R)$ shown in Figure~\ref{sigma}.
From the condition $\sigma(R)\sim 1+z$ it follows that $R(z)\sim 80\,h^{-1}\times 10^{-(1+z)^{0.6}}$\,Mpc and substitution into~(\ref{uz1}) gives
\begin{eqnarray}\label{uz2}
u(z)\,\sim\, 0.01\sqrt{1+z}~10^{-(1+z)^{0.6}}\,.
\end{eqnarray}
The function~(\ref{uz2}) is plotted in Figure~\ref{uplot}. It may be seen that $u(z)$ has a maximum of $\sim 10^{-3}$ at $z=0$ and decreases as $z$ increases. The ratio of the string density to the background string density was $u\sim 10^{-4}$ for galaxy-scale fluctuations becoming nonlinear at $z\sim 3$ and the density ratio for the first stars forming at $z\sim 10$ was $u\sim 2\times 10^{-6}$.

Consider the formation of a galaxy from an overdense region that separated from the Hubble flow at $z\sim 3$ 
with $u\sim 10^{-4}$. At the edge of a galaxy of mass $M$ and radius $R_g$, equations~(\ref{urs}) and~(\ref{u3}) both give $u\sim (M/R_g^2)(M_U/R_U^2)^{-1}$. It follows from equation~(\ref{rg2}) that $u\sim 1$ for a galaxy at the present time. At redshift $z$, $M_U\propto \overline\rho(z)R_U^3$, where $\overline\rho(z)\propto(1+z)^3$ is the average matter density at redshift $z$, and $R_U\propto (1+z)^{-3/2}$ since $R_U\propto t$ and $a(t)=1/(1+z)\propto t^{2/3}$ in the matter era. Thus $M_U/R_U^2\propto (1+z)^3R_U\propto (1+z)^{3/2}$ and it follows that $M_U/R_U^2$ decreases by a factor of order $8$ from $z\sim 3$ to the present time. The increase in $u$ from $u\sim 10^{-4}$ at the time of separation from the Hubble flow to $u\sim 1$ at the present time therefore implies that the radius $R_g$ decreases by a factor of order $(8\times 10^{-4})^{-1/2}\sim 35$ during the process of galaxy formation.

 \section{Calculation of the additional acceleration}\label{appcalc}
\subsection{Direct calculation of the force on the centre}\label{mforce}
\subsubsection{The force on a test mass due to its Machian strings}
Consider the Machian strings around a test mass $m$ in the presence of a mass $M$ and let the direction of a Machian string be specified by spherical polar coordinates $\theta$ and $\phi$, where the $z$ axis is along the line of centres joining the two masses. Since the distribution of Machian strings connected to $m$ is uniform at large distances, the total force acting on the centre of $m$ is given by
\begin{eqnarray}\label{f1}
 {\bf F}\,=\, \frac{1}{4\pi}\int {\bf T}(\theta_f,\phi_f)\sin\theta_f\,d\theta_f d\phi_f\,,
\end{eqnarray}
where ${\bf T}(\theta_f,\phi_f)$ is the tension at the centre of $m$ in a string whose direction at large distances has polar coordinates $\theta_f$ and $\phi_f$. The magnitude of the string tension at $m$ is $T(R)[1-f(u_0)]$ for all strings, from~(\ref{tm0}), where $u_0$ is the value of $u$ at $m$. The component of the total force along the line of centres, from $M$ to $m$, is 
\begin{eqnarray}\label{f2}
 F\,=\, -\frac{T(R)}{4\pi}[1-f(u_0)]\iint\cos\theta_i(\theta_f,\phi_f)\,\sin\theta_f d\theta_f d\phi_f\,,\hspace*{-0.5cm}\nonumber\\ 
\end{eqnarray}
 where $\theta_i$ is the angle between the string direction at the centre of $m$ and the line of centres. 

The maximum additional acceleration occurs in the limit that all the strings connected to the test mass have the maximum tension $T(R)$ and are all aligned along the same direction at the centre of $m$. Since $T(R_i)=Gmm_i/R_i^2$ for a string of length $R_i$ connected to a distant mass $m_i$, the maximum additional acceleration on a test mass is 
\begin{eqnarray}\label{amax}
 a_{max}\,=\,\sum_i\frac{Gm_i}{R_i^2}\sim\int\frac{G\,\overline\rho\,dV}{r^2}\,=\,4\pi G\,\overline\rho R_U\,,
\end{eqnarray}
where $\overline\rho$ is the average matter density at the present time. The mass of the observable Universe, $M_U$, is given by 
\begin{eqnarray}\label{mobs}
M_U\,=\,\frac{4\pi}{3}\overline\rho R_U^3\,,
\end{eqnarray}
so the maximum acceleration may be written in the form
\begin{eqnarray}
a_{max} \,=\,3\Big(\frac{GM_U}{R_Uc^2}\Big)\frac{c^2}{R_U}\,.
\end{eqnarray}
The acceleration due to the Machian strings corresponding to the force~(\ref{f2}) is therefore
\begin{eqnarray}\label{af}
 a\,=\, -3\Big(\frac{GM_U}{R_Uc^2}\Big)\frac{c^2}{R_U}[1-f(u_0)]\,\eta\,,
\end{eqnarray}
where $\eta$ is the asymmetry in the strings defined by
\begin{eqnarray}\label{asym}
\eta\,=\,\frac{1}{4\pi}\iint\cos\theta_i(\theta_f,\phi_f)\,\sin\theta_f d\theta_f d\phi_f\,.
\end{eqnarray}

\subsubsection{Calculation of the asymmetry}
To calculate the asymmetry~(\ref{asym}) it is first necessary to calculate the paths of Machian strings connected to the test mass. The string paths may be calculated by numerical integration of equation~(\ref{curv}). Recalling that ${\mathcal E}=T(R)[1+f(u)]$ and ${\mathcal E}+I\approx T(R)[-1+f(u)]$, from equation~(\ref{dens}) and equations~(\ref{tens}) and~(\ref{tm}), it follows that ${\pmb\nabla}{\mathcal E}/({\mathcal E}+I)= f^\prime(u){\pmb\nabla}u/(-1+f)$.  The tangent vector to the string path is ${\bf x}^\prime={\bf e}_r=(\sin\theta\cos\phi,\sin\theta\sin\phi,\cos\theta)$ so ${\bf x}^{\prime\prime}=\theta^\prime\,{\bf e}_\theta + \sin\theta\,\phi^\prime\,{\bf e}_\phi$. Taking ${\bf m}={\bf e}_\theta$ and ${\bf m}={\bf e}_\phi$ in~(\ref{curv}) then gives the string path equations
\begin{eqnarray}\label{et}
&&\hspace*{-0.6cm} \theta^\prime\,=\,(\cos\theta\cos\phi,\cos\theta\sin\phi,-\sin\theta).\frac{f^\prime(u){\pmb\nabla}u}{-1+f}\hspace*{0.9cm} \\ \mbox{and}\hspace*{0.2cm} &&\sin\theta\,\phi^\prime\,=\,(-\sin\phi,\,\cos\phi,\,0).\frac{f^\prime(u){\pmb\nabla}u}{-1+f}\,.\hspace*{0.7cm}\label{ep}
\end{eqnarray}

Equations~(\ref{et}) and~(\ref{ep}) may be integrated outwards from the centre of the test mass $m$ to find the polar coordinates $\theta_f$ and $\phi_f$ for the final string direction as a function of the initial polar coordinates $\theta_i$ and $\phi_i$. The values of $\cos\theta_i$ for given values of $\theta_f$ and $\phi_f$ may be found by interpolation and the asymmetry~(\ref{asym}) may then be evaluated numerically.


\subsubsection{String path equations for spherical and cylindrical string densities around M}
When the strings of the mass $M$ have a spherical distribution, the density ratio $u$ at the centre of $m$ is given by equation~(\ref{u1}), where $\rho$ is the distance between the two masses in units of $R_U\sqrt{M/M_U}$. Equations~(\ref{et}) and~(\ref{ep}) require the density ratio at a general point along one of the strings of $m$. If ${\bf x}=(x,y,z)$ is the position relative to the centre of $m$, in units of $R_U\sqrt{M/M_U}$, the density ratio at ${\bf x}$ is given by $u({\bf x})=[x^2+y^2+(\rho-z)^2]^{-1}$. Equations~(\ref{et}) and~(\ref{ep}) then become
\begin{eqnarray}\label{et2}
&&\hspace*{-1.3cm}\theta^\prime=\frac{2uf^\prime(u)}{1-f}\Big(\frac{x\cos\theta\cos\phi+y\cos\theta\sin\phi+(\rho-z)\sin\theta}{x^2+y^2+(\rho-z)^2}\Big)\hspace*{-0.5cm}\nonumber\\ && \\ \hspace*{-0.2cm}\mbox{and}\hspace*{0.3cm} &&\sin\theta\,\phi^\prime\,=\,\frac{2uf^\prime(u)}{1-f}\Big(\frac{-x\sin\phi+y\cos\phi}{x^2+y^2+(\rho-z)^2}\Big)\,.\nonumber\\ && \label{ep2}
\end{eqnarray}

Now consider a galaxy of mass $M$ with strings in a cylindrical distribution and consider a test mass $m$ in the plane of a galaxy. The $z$ axis is along the line of centres joining the two masses so let the normal to the plane of the galaxy be along the $y$ axis. A point with position vector ${\bf x}$ relative to the centre of $m$ is at a distance $\sqrt{x^2+(\rho-z)^2}$ from the centre of $M$ in the plane of the galaxy and at a distance $y$ in the direction normal to the plane of the galaxy so, according to equation~(\ref{u4}), the string density ratio at 
${\bf x}$ is $u({\bf x})=2\widetilde\lambda[\pi\sqrt{x^2+(\rho-z)^2}(y^2+\widetilde\lambda^2)]^{-1}$. Equations~(\ref{et}) and~(\ref{ep}) then give
\begin{eqnarray}\label{et3}
&&\hspace*{-0.5cm}\theta^\prime\,=\,\frac{uf^\prime(u)}{1-f}\Big(\frac{x\cos\theta\cos\phi-(z-\rho)\sin\theta}{x^2+(\rho-z)^2}\nonumber\\
&&\hspace*{3.2cm}+~\frac{2y\cos\theta\sin\phi}{y^2+\widetilde\lambda^2}\Big)\\\mbox{and}\hspace*{0.3cm} &&\sin\theta\,\phi^\prime\,=\,\frac{uf^\prime(u)}{1-f}\Big(\frac{-x\sin\phi}{x^2+(\rho-z)^2}+\frac{2y\cos\phi}{y^2+\widetilde\lambda^2}\Big)\,.\nonumber\\\label{ep3}
\end{eqnarray}

\subsection{Calculation using interaction energy for $\rho\gg 1$}\label{appecalc}
The change in the total energy of the Machian strings of $m$ corresponding to the interaction with the Machian strings of $M$ defined by equation~(\ref{dens}) is
\begin{eqnarray}\label{change}
 \Delta E\,=\,\sum_i\frac{Gm\overline m_i}{R_i^2}\int_0^{R_i} \!f[u({\bf x}-{\bf x}_M)]\,ds\,,
\end{eqnarray}
 where $s$ denotes the path length along the $i^{th}$ string, ${\bf x}$ is the position vector of a point on one of the strings of $m$ relative to the centre of $m$, ${\bf x}_M$ is the position vector of the centre of $M$ relative to the centre of $m$ and $s= |{\bf x}|$. When the basic postulate of the string model stated in Section~\ref{msm} is taken into account, the corresponding change in energy of the whole system, including all the other masses other than $m$ and $M$, is equal to $-\Delta E$. The additional acceleration of the mass $m$ due to the interaction between the Machian strings is therefore given by
\begin{eqnarray}\label{a1}
 a\,=\,\frac{1}{m}\frac{d\,\Delta E}{dr}\,.
\end{eqnarray}

When $\rho\gg 1$, where $\rho$ is the distance between the centres of the two masses in units of $R_U\sqrt{M/M_U}$,  
the asymmetry in the Machian strings of the mass $m$ may be neglected and the length $R_i$ of the string connecting the mass $m$ to the mass $\overline m$ is simply the distance between the centres. The total mass $\overline m$ in an elemental solid angle $d\Omega$ and thickness $dR$ at radius $R$ is $\overline \rho R^2 dRd\Omega$, where $\overline\rho$ is the average matter density and spherical polar coordinates are defined relative to the centre of $m$. After integrating over the azimuthal angle, equation~(\ref{change}) becomes
\begin{eqnarray}\label{change2}
 \Delta E\,\approx\,2\pi Gm\overline\rho\int_0^{R_U}\!\!\!dR\!\int_0^\pi \!\!\sin\theta\,d\theta \int_0^R\!\!f[u({\bf x}-{\bf x}_M)]\,ds\,.\hspace*{-0.5cm}\nonumber\\
\end{eqnarray}

Since $f\rightarrow 0$ as $s\rightarrow \infty$, the integral over $s$ in~(\ref{change2}) is insensitive to the value of $R$, so $R$ can be replaced by $R_U$ in the upper limit of the integral over $s$ and the integral over $R$ then simply gives a factor of $R_U$. Substituting for $\overline\rho$ using~(\ref{mobs}) then gives
\begin{eqnarray}\label{int}
 \Delta E\,\approx\,
\frac{3mc^2}{2R_U}\Big(\frac{GM_U}{R_Uc^2}\Big)\!\int_0^{R_U}\!\!\!\!ds\!\int_0^\pi \!\!\!d\theta\,\sin\theta \,
f[u({\bf x}-{\bf x}_M)]\,.\hspace*{-0.5cm}\nonumber\\
\end{eqnarray}
The volume element is $d^3{\bf x}=s^2\sin\theta\, dsd\theta d\phi$ so~(\ref{int}) can be written as
\begin{eqnarray}\label{int2}
 \Delta E\,\approx\,\frac{3mc^2}{4\pi R_U}\Big(\frac{GM_U}{R_Uc^2}\Big)\int\frac{d^3{\bf x}}{|{\bf x}|^2}
 \,f[u({\bf x}-{\bf x}_M)]\,.~~
\end{eqnarray}
To evaluate $\Delta E$ it is convenient to change the origin so that ${\bf x}$ is now the position vector relative to the centre of $M$ and equation~(\ref{int2}) then becomes
\begin{eqnarray}\label{int3}
 \Delta E\,\approx\,\frac{3mc^2}{4\pi R_U}\Big(\frac{GM_U}{R_Uc^2}\Big)\int\frac{d^3{\bf x}}{|{\bf x}+{\bf x}_M|^2}
 \,f[u({\bf x})]\,.~~
\end{eqnarray}

\subsubsection{Spherical distribution of strings around $M$, $\rho\gg 1$}
 If x denotes the distance $|{\bf x}|$ in units of $R_U\sqrt{M/M_U}$ then, since 
$|{\bf x}_M|=\rho R_U\sqrt{M/M_U}$, equation~(\ref{int3}) becomes 
\begin{eqnarray}\label{int4}
 &&\hspace*{-1cm}\Delta E\,\approx\, \frac{3mc^2}{2}\sqrt{\frac{M}{M_U}}\Big(\frac{GM_U}{R_Uc^2}\Big)\!\int_0^{\sqrt{\frac{M_U}{M}}}\nonumber\\ && \hspace*{2cm}\int_0^\pi \!\!\!\frac{x^2dx\sin\theta d\theta}{x^2 + 2\rho x\cos\theta + \rho^2}\,f[u(x)]\,,\nonumber\\
\end{eqnarray}
using spherical polar coordinates centred on $M$. For a spherical string distribution, $u=1/x^2$ and $f(u)\approx \gamma/x^2$ when $x\gg 1$. Equation~(\ref{int4}) may then be integrated with respect to $x$, using $\int\,dx/[(x+a)^2+b^2]=b^{-1}\tan^{-1}[(x+a)/b]$, to give
\begin{eqnarray}
&& \hspace*{-1.3cm}\Delta E\,\approx\,\frac{3\gamma mc^2}{2\rho}\sqrt{\frac{M}{M_U}}\Big(\frac{GM_U}{R_Uc^2}\Big)\int_0^\pi \!d\theta\,\Big[\theta-\nonumber\\
&& \hspace*{1.5cm}\cot^{-1}\Big(\frac{\sqrt{M_U/M}+\rho\cos\theta}{\rho\sin\theta}\Big)\Big]\,.~
\end{eqnarray}
Using the expansion $\cot^{-1}z=1/z-1/3z^3+\dots$ we find
\begin{eqnarray}
\Delta E\,\approx\,\frac{3\gamma mc^2}{2\rho}\sqrt{\frac{M}{M_U}}\Big(\frac{GM_U}{R_Uc^2}\Big)\Big[\frac{\pi^2}{2}-2\rho\sqrt{\frac{M}{M_U}}+\dots\Big]\,,\hspace*{-0.5cm}\nonumber\\
\end{eqnarray}
for $\rho\ll \sqrt{M_U/M}$, and the corresponding acceleration~(\ref{a1}) is
\begin{eqnarray}\label{accs}
a\,\approx\,\frac{3\pi^2\gamma}{4\rho^2}\Big(\frac{GM_U}{R_Uc^2}\Big)\frac{c^2}{R_U}\,.
\end{eqnarray}

\subsubsection{Cylindrical distribution of strings around $M$, $\rho\gg 1$}
For a cylindrically symmetric distribution of strings around a galaxy, the interaction energy~(\ref{int3}) for a test particle in the plane of the galaxy may be evaluated using cylindrical polar coordinates centred on $M$ to give
\begin{eqnarray}\label{eint}
&&\hspace*{-1.3cm}\Delta E \approx \frac{3mc^2}{4\pi R_U}\Big(\frac{GM_U}{R_Uc^2}\Big)\sqrt{\frac{M}{M_U}}
\int_0^{2\pi}\!\!d\theta\nonumber\\
&&\hspace*{-0.6cm}\int_0^{\sqrt{M_U/M}}\!\!\!\!dx \int_{-\infty}^\infty\frac{x\,f[u({\bf x})]\,dz}{x^2-2\rho x\cos\theta+\rho^2+z^2}\,,
\end{eqnarray}
where the normal to the plane of the galaxy is now along the $z$ axis and distances are again in units of $R_U\sqrt{M/M_U}$. The required generalisation of equation~(\ref{u4}) at a point in the plane of the galaxy is $u({\bf x})=2\widetilde\lambda[\pi x(\widetilde\lambda^2+z^2)]^{-1}$ and $f(u)\approx\gamma u$ for $x\gg 1$. The corresponding additional acceleration~(\ref{a1}), in units of $c^2/R_U$, is
\begin{eqnarray}\label{inv3}
&& \hspace*{-1.2cm}\frac{d}{d\rho}\left\{\frac{3GM_U}{4\pi R_Uc^2}\int_0^{2\pi}\!\!d\theta\int_0^{\sqrt{M_U/M}}\!\!dx\right.\nonumber\\
&& \left.\int\!\frac{2\gamma\widetilde\lambda~dz}{\pi(\widetilde\lambda^2+z^2)(x^2-2\rho x\cos\theta+\rho^2+z^2)}\right\}\nonumber\\
&&\hspace*{-1cm} =~\frac{d}{d\rho}\left\{\frac{3GM_U}{4\pi R_Uc^2}\int_0^{2\pi}\!\!d\theta\int_0^{\sqrt{M_U/M}}\!\!\frac{2\gamma\,dx}{d(\widetilde	\lambda+d)}\right\},
\end{eqnarray}
where $d=\sqrt{x^2-2\rho x\cos\theta+\rho^2}$ and use has been made of the integral $\int_{-\infty}^\infty dz/[(a^2+z^2)(b^2+z^2)]=\pi[ab(a+b)]^{-1}$.

\subsection{Numerical results}
\subsubsection{Spherically symmetric distribution of strings around $M$}
 The additional acceleration for a spherically symmetric string distribution and an interaction function with $\alpha=0.68$ and $\beta=1$ was calculated using the method described in Section~\ref{mforce} and the result is shown in Figure~\ref{spherical}. The result for $\rho\gg 1$ is in good agreement with the formula~(\ref{accs}) derived using the interaction energy method described in Section~\ref{appecalc}.

The additional acceleration in the limit $\rho\ll 1$ can be understood by considering the formula~(\ref{af}). The value $\alpha=0.68$ corresponds to $A=1$, by equation~(\ref{aa}). Numerical calculations show that, when $A=1$, the asymmetry in the strings tends to $1$ as $\rho\rightarrow 0$. Substituting $1-f(u)\approx 1/bu^2 =\rho^4/b$ for $\rho\ll 1$ into~(\ref{af}) then gives equation~(\ref{limit}). 

The suppression of dark matter effects as $\rho\rightarrow 0$ is due to the fact that the interaction function $f(u)$ in~(\ref{intf2}) tends a constant when the Newtonian gravitational acceleration is much larger than $a_0$ and occurs when $A<1$ as well as when $A=1$. When $A<1$, the string tension tends to a constant proportional to $1-A$ and numerical calculations show that the asymmetry tends to zero as $\rho\rightarrow 0$ in such a way that the product of the string tension and the asymmetry is still proportional to $\rho^4$.

\subsubsection{Cylindrically symmetric distribution of strings around $M$}
The additional acceleration acting on a test particle for a galaxy whose strings have a cylindrical density distribution, with density given by equation~(\ref{u4}), was calculated using the method described in Section~\ref{mforce}. The additional acceleration and the corresponding velocity rotation curve were calculated for various values of the parameters $\alpha$ and $\lambda$ with $\beta=1$. The values $\alpha=0.68$ and $\lambda=0.5$ were found to give a very close fit to the MOND velocity rotation curve, as discussed in Section~\ref{form}. 

\begin{figure}[h]
\hspace*{-3.5cm}
\includegraphics[height=6cm,width=5cm]{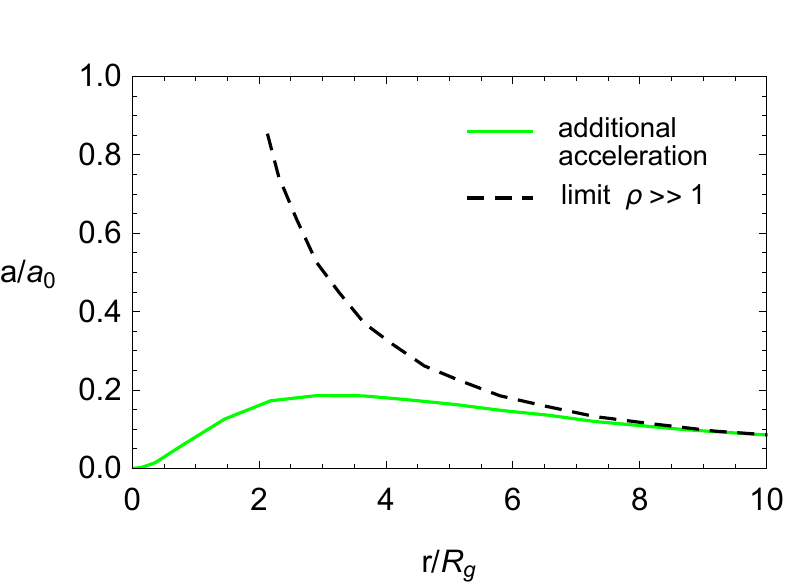}
\caption{\label{ctotal10c} The additional acceleration shown in Figure~\ref{ctotal} for a cylindrically symmetric string distribution around a mass $M$ with $\lambda=0.5$, $\alpha=0.68$ and $\beta=1$, calculated using the method described in Section~\ref{mforce}. The dashed line shows the limit for $\rho\gg 1$ calculated using the interaction energy method described in Section~\ref{appecalc}.}
\end{figure}

The additional acceleration shown in Figure~\ref{ctotal} can be checked for $\rho\gg 1$ using the interaction energy method described in Section~\ref{appecalc}. Numerical evaluation of equation~(\ref{inv3}) gives the dashed curve shown in Figure~\ref{ctotal10c}. The agreement between the two methods for $\rho\gg 1$ confirms the fit to the MOND rotation curve shown in Figure~\ref{rotation}.

\begin{figure}[h]
\hspace*{-3.5cm}
\includegraphics[height=6cm,width=5cm]{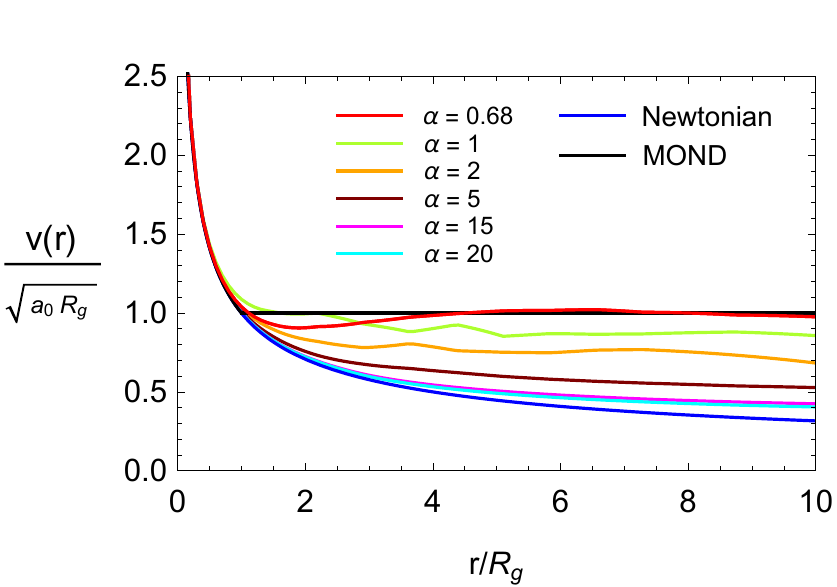}
\caption{\label{figalpha} The effect on the rotation curve of varying the parameter $\alpha$, with $\lambda=0.5$. The curve with $\alpha=0.68$ is shown in Figure~\ref{rotation}.}
\end{figure}

\begin{figure}[h]
\hspace*{-3.5cm}
\includegraphics[height=6cm,width=5cm]{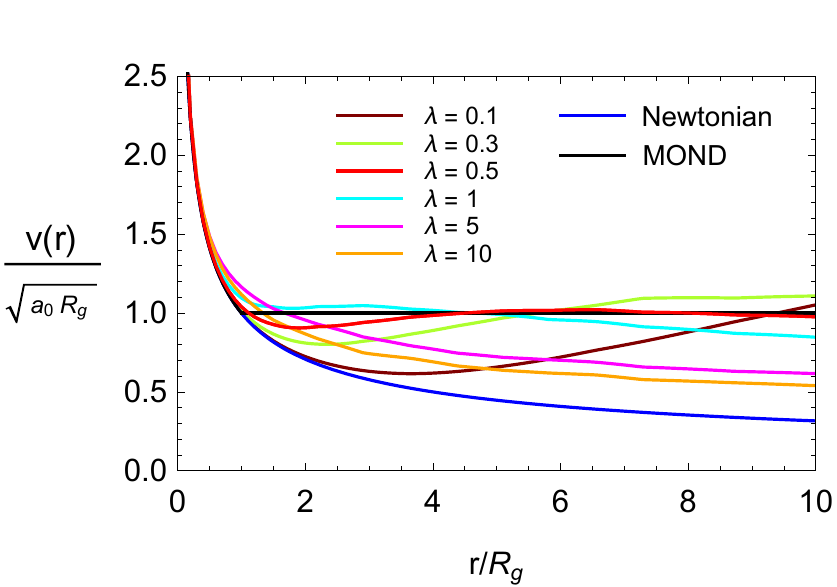}
\caption{\label{figlambda} The effect on the rotation curve of varying the parameter $\lambda$, with $\alpha=0.68$. The curve with $\lambda=0.5$ is shown in Figure~\ref{rotation}.}
\end{figure}

It is of interest to investigate the sensitivity of the velocity rotation curve to changes in the values of $\alpha$ and $\lambda$. The rotation curves for different values of $\alpha$, with $\lambda=0.5$, are shown in Figure~\ref{figalpha}. In the limit $\alpha\rightarrow\infty$, the interaction function $f(u)$ tends to a constant and the additional acceleration tends to zero with the result that the total acceleration is equal to the Newtonian acceleration. Figure~\ref{figlambda} shows the rotation curves for different values of $\lambda$ with $\alpha=0.68$.  The value $\alpha=0.68$ corresponds to $A=1$, by equation~(\ref{aa}), which is the largest value of $A$ consistent with the stability condition $f(u)<1$ derived in Appendix~\ref{appmodel}. 

\end{appendix}

\bibliography{master4}

\begin{thebibliography}{21}
\expandafter\ifx\csname natexlab\endcsname\relax\def\natexlab#1{#1}\fi
\expandafter\ifx\csname bibnamefont\endcsname\relax
  \def\bibnamefont#1{#1}\fi
\expandafter\ifx\csname bibfnamefont\endcsname\relax
  \def\bibfnamefont#1{#1}\fi
\expandafter\ifx\csname citenamefont\endcsname\relax
  \def\citenamefont#1{#1}\fi
\expandafter\ifx\csname url\endcsname\relax
  \def\url#1{\texttt{#1}}\fi
\expandafter\ifx\csname urlprefix\endcsname\relax\def\urlprefix{URL }\fi
\providecommand{\bibinfo}[2]{#2}
\providecommand{\eprint}[2][]{\url{#2}}

\bibitem[{\citenamefont{Longair}(1998{\natexlab{a}})}]{longair}
\bibinfo{author}{\bibfnamefont{M.~S.} \bibnamefont{Longair}},
  \emph{\bibinfo{title}{Galaxy Formation}} (\bibinfo{publisher}{Springer},
  \bibinfo{year}{1998}{\natexlab{a}}), \bibinfo{note}{chapter 12}.

\bibitem[{\citenamefont{Zwicky}(1937)}]{zwicky}
\bibinfo{author}{\bibfnamefont{F.}~\bibnamefont{Zwicky}}, \bibinfo{journal}{Ap.
  J.} \textbf{\bibinfo{volume}{86}}, \bibinfo{pages}{217}
  (\bibinfo{year}{1937}).

\bibitem[{\citenamefont{Rubin et~al.}(1978)\citenamefont{Rubin, Ford, and
  Thonnard}}]{rubin}
\bibinfo{author}{\bibfnamefont{V.~C.} \bibnamefont{Rubin}},
  \bibinfo{author}{\bibfnamefont{W.~K.} \bibnamefont{Ford}}, \bibnamefont{and}
  \bibinfo{author}{\bibfnamefont{N.}~\bibnamefont{Thonnard}},
  \bibinfo{journal}{Ap. J.} \textbf{\bibinfo{volume}{225}},
  \bibinfo{pages}{L107} (\bibinfo{year}{1978}).

\bibitem[{\citenamefont{Milgrom}(1983)}]{milgrom}
\bibinfo{author}{\bibfnamefont{M.}~\bibnamefont{Milgrom}},
  \bibinfo{journal}{Astrophys. J.} \textbf{\bibinfo{volume}{270}},
  \bibinfo{pages}{365} (\bibinfo{year}{1983}).

\bibitem[{\citenamefont{Bekenstein}(2004)}]{teves}
\bibinfo{author}{\bibfnamefont{J.~D.} \bibnamefont{Bekenstein}},
  \bibinfo{journal}{PRD} \textbf{\bibinfo{volume}{70}}, \bibinfo{pages}{083509}
  (\bibinfo{year}{2004}), \bibinfo{note}{ast-ph/0410182}.

\bibitem[{\citenamefont{Moffat}(2006)}]{mog}
\bibinfo{author}{\bibfnamefont{J.~W.} \bibnamefont{Moffat}},
  \bibinfo{journal}{J. Cosmol. Astropart. Phys.} \textbf{\bibinfo{volume}{3}},
  \bibinfo{pages}{4} (\bibinfo{year}{2006}), \bibinfo{note}{gr-gc/0506021}.

\bibitem[{\citenamefont{B{\"o}hmer et~al.}(2008)\citenamefont{B{\"o}hmer,
  Harko, and Lobo}}]{boehmer}
\bibinfo{author}{\bibfnamefont{C.~G.} \bibnamefont{B{\"o}hmer}},
  \bibinfo{author}{\bibfnamefont{T.}~\bibnamefont{Harko}}, \bibnamefont{and}
  \bibinfo{author}{\bibfnamefont{F.~S.~N.} \bibnamefont{Lobo}},
  \bibinfo{journal}{Astropart. Phys.} \textbf{\bibinfo{volume}{29}},
  \bibinfo{pages}{386} (\bibinfo{year}{2008}), \bibinfo{note}{gr-qc/0709.0046}.

\bibitem[{\citenamefont{Mannheim}(2006)}]{mannheim2}
\bibinfo{author}{\bibfnamefont{P.~D.} \bibnamefont{Mannheim}},
  \bibinfo{journal}{Prog. Part. Nucl. Phys.} \textbf{\bibinfo{volume}{56}},
  \bibinfo{pages}{340} (\bibinfo{year}{2006}), \bibinfo{note}{ast-ph/0505266}.

\bibitem[{\citenamefont{Essex}(2016)}]{paper3}
\bibinfo{author}{\bibfnamefont{D.~W.} \bibnamefont{Essex}}
  (\bibinfo{year}{2016}), \bibinfo{note}{arXiv:1608.06840}.

\bibitem[{\citenamefont{McGaugh}(2004)}]{mcgaugh}
\bibinfo{author}{\bibfnamefont{S.~S.} \bibnamefont{McGaugh}},
  \bibinfo{journal}{Ap. J.} \textbf{\bibinfo{volume}{611}}, \bibinfo{pages}{26}
  (\bibinfo{year}{2004}), \bibinfo{note}{table 2}.

\bibitem[{\citenamefont{Sanders}(1990)}]{sanders2}
\bibinfo{author}{\bibfnamefont{R.~H.} \bibnamefont{Sanders}},
  \bibinfo{journal}{Astron. Astrophys. Rev.} \textbf{\bibinfo{volume}{2}},
  \bibinfo{pages}{1} (\bibinfo{year}{1990}).

\bibitem[{\citenamefont{Bennett et~al.}(2012)}]{bennett}
\bibinfo{author}{\bibfnamefont{C.~L.} \bibnamefont{Bennett}}
  \bibnamefont{et~al.}, p. \bibinfo{pages}{129} (\bibinfo{year}{2012}),
  \bibinfo{note}{arXiv:1212.5225}.

\bibitem[{\citenamefont{Longair}(1998{\natexlab{b}})}]{longair2}
\bibinfo{author}{\bibfnamefont{M.~S.} \bibnamefont{Longair}},
  \emph{\bibinfo{title}{Galaxy Formation}} (\bibinfo{publisher}{Springer},
  \bibinfo{year}{1998}{\natexlab{b}}), \bibinfo{note}{chapter 16}.

\bibitem[{\citenamefont{Fall and Efstathiou}(1980)}]{fall}
\bibinfo{author}{\bibfnamefont{S.~M.} \bibnamefont{Fall}} \bibnamefont{and}
  \bibinfo{author}{\bibfnamefont{G.}~\bibnamefont{Efstathiou}},
  \bibinfo{journal}{MNRAS} \textbf{\bibinfo{volume}{193}}, \bibinfo{pages}{189}
  (\bibinfo{year}{1980}).

\bibitem[{\citenamefont{{Wolfram Research, Inc.}}()}]{mathematica}
\bibinfo{author}{\bibnamefont{{Wolfram Research, Inc.}}},
  \emph{\bibinfo{title}{Mathematica 10.2}},
  \urlprefix\url{https://www.wolfram.com}.

\bibitem[{\citenamefont{Essex}(2015)}]{paper2n}
\bibinfo{author}{\bibfnamefont{D.~W.} \bibnamefont{Essex}}
  (\bibinfo{year}{2015}), \bibinfo{note}{arXiv:1507.02914. Note that the
  present paper uses a simpler interaction function $f(u)$.}

\bibitem[{\citenamefont{Clowe et~al.}(2006)}]{clowe}
\bibinfo{author}{\bibfnamefont{D.}~\bibnamefont{Clowe}} \bibnamefont{et~al.},
  \bibinfo{journal}{Ap. J. Lett.} \textbf{\bibinfo{volume}{648}},
  \bibinfo{pages}{L109} (\bibinfo{year}{2006}),
  \bibinfo{note}{astro-ph/0608407}.

\bibitem[{\citenamefont{Jee et~al.}(2012)}]{jee}
\bibinfo{author}{\bibfnamefont{M.~J.} \bibnamefont{Jee}} \bibnamefont{et~al.},
  \bibinfo{journal}{Ap. J.} \textbf{\bibinfo{volume}{747}}, \bibinfo{pages}{96}
  (\bibinfo{year}{2012}), \bibinfo{note}{arXiv:1202.6368}.

\bibitem[{\citenamefont{Govoni and Feretti}(2004)}]{govoni}
\bibinfo{author}{\bibfnamefont{F.}~\bibnamefont{Govoni}} \bibnamefont{and}
  \bibinfo{author}{\bibfnamefont{L.}~\bibnamefont{Feretti}},
  \bibinfo{journal}{Int. J. Mod. Phys. D} \textbf{\bibinfo{volume}{13}},
  \bibinfo{pages}{1549} (\bibinfo{year}{2004}), \bibinfo{note}{ast-ph/0403694}.

\bibitem[{\citenamefont{Pitjev and Pitjeva}(2013)}]{pitjev}
\bibinfo{author}{\bibfnamefont{N.~P.} \bibnamefont{Pitjev}} \bibnamefont{and}
  \bibinfo{author}{\bibfnamefont{E.~V.} \bibnamefont{Pitjeva}}
  (\bibinfo{year}{2013}), \bibinfo{note}{arXiv:1306.5534}.

\bibitem[{\citenamefont{Zentner}(2007)}]{zentner}
\bibinfo{author}{\bibfnamefont{A.~R.} \bibnamefont{Zentner}},
  \bibinfo{journal}{Int. J. Mod. Phys. D} \textbf{\bibinfo{volume}{16}},
  \bibinfo{pages}{763} (\bibinfo{year}{2007}), \bibinfo{note}{ast-ph/0611454,
  Figure 1}.

\end{thebibliography}

\end{document}